\documentclass[aps,prl,twocolumn,showpacs,superscriptaddress,groupedaddress]{revtex4}  

\usepackage{graphicx}  
\usepackage{amssymb}   
\usepackage{amsmath}
\usepackage{hyperref}
\usepackage{xcolor}

\usepackage{endnotes}
\let\footnote=\endnote

\begin{document}

\begin{flushleft}
\mbox{HRI-RECAPP-2014-015}
\end{flushleft}

\title{Drell-Yan Production at Threshold to Third Order in QCD}

\author{Taushif Ahmed\footnote{taushif@hri.res.in}}
\affiliation{Regional Centre for Accelerator-based Particle Physics,
Harish-Chandra Research Institute,  
Allahabad, India}

\author{Maguni Mahakhud\footnote{maguni@hri.res.in}}
\affiliation{Regional Centre for Accelerator-based Particle Physics,
Harish-Chandra Research Institute,  
Allahabad, India}

\author{Narayan Rana\footnote{narayan@hri.res.in}}
\affiliation{Regional Centre for Accelerator-based Particle Physics,
Harish-Chandra Research Institute,  
Allahabad, India}

\author{V. Ravindran\footnote{ravindra@imsc.res.in}}
\affiliation{The Institute of Mathematical Sciences, Chennai, India }

\date{\today}

\begin{abstract}
The recent computation on the full threshold contributions to Higgs boson production at next-to-next-to-next-to-leading order (N$^3$LO) in QCD
contains valuable information on the soft gluons resulting from virtual and real
emission partonic subprocesses.  We use those from the real emissions 
to obtain the corresponding soft
gluon contributions to Drell-Yan production and determine the missing $\delta(1-z)$ part
of the  N$^3$LO.  
The numerical impact of threshold effects demonstrates the importance
of our results in the precision study with the Drell-Yan process at the LHC.
\end{abstract}

\maketitle
Discovery of the Higgs boson and the exclusion limits on its mass strongly depend on
the precise knowledge of its production mechanism in the standard model (SM) and its extensions.
The fixed as well as resummed next to next to leading order (NNLO) \cite{nnlo} and leading log (NNLL) \cite{nnll}
quantum chromodynamics (QCD) corrections
supplemented with two-loop electroweak effects \cite{ewnnlo} played an important role in the discovery
of Higgs boson by ATLAS and CMS collaborations \cite{AtlasCMS} at the Large Hadron Collider.
Also for Drell-Yan (DY) production, NNLO \cite{dynnlo} and NNLL \cite{Vogt:2000ci} QCD results are known.
There have been several attempts to go beyond the NNLO level in QCD.  The gluon and quark form
factors \cite{3lffmoch, Baikov:2009bg, Gehrmann:2010ue}, the mass factorization kernels \cite{Moch:2004pa}
and the renormalization constant \cite{Chetyrkin:1997un} for the effective operator describing the coupling of the 
Higgs boson with the SM fields in the infinite top quark mass limit up to the three-loop level in dimensional regularization, 
with space-time dimension $d = 4 + \epsilon$, enabled one to obtain
the next-to-next-to-next-to-leading order (N$^3$LO) threshold effects \cite{n3losv, n3losvRavi}, often called soft plus virtual (SV) contributions, to the
inclusive Higgs boson and DY productions at the LHC , excluding the term proportional to $\delta(1-z)$ where
the scaling parameter $z=m_H^2/\hat s$ for the Higgs boson and $z=m_{l^+l^-}^2/\hat s$ for DY.  Here $m_H$, $m_{l^+l^-}$, and $\hat s$ are the mass of the Higgs boson,
invariant mass of the dileptons and center of mass energy of the partonic reaction responsible for production mechanism, respectively.
Note that the finite mass factorized threshold contribution to the inclusive production
cross section is expanded in terms of $\delta(1-z)$ and ${\cal D}_i(z)$, where
\begin{eqnarray}
{\cal D}_i(z) = \left({\ln^{i}(1-z) \over (1-z)}\right)_+ \, .
\end{eqnarray}
The $\delta(1-z)$ part of N$^3$LO threshold contribution was not known until recently because 
of the lack of information on the complete soft contributions coming from  
real radiation processes, while the exact two- \cite{Gehrmann:2005pd} and three-loop \cite{Gehrmann:2010ue} quark and gluon form factors 
and NNLO soft contributions \cite{deFlorian:2012za} to all orders in $\epsilon$ are already known.  
The full threshold N$^3$LO result for Higgs boson production has now become reality due the recent 
computation by Anastasiou \textit{et al.} \cite{Anastasiou:2014vaa} who have computed all these soft effects 
resulting from gluon radiations, which constitute the missing part.       
In this Letter, we investigate the impact of these soft gluon contributions   
on the $\delta(1-z)$ part of the N$^3$LO to DY production.  
This completes the full N$^3$LO threshold contribution to DY production.

The production cross section of a heavy particle, namely, a Higgs boson or a pair of leptons at the hadron colliders
can be computed using
\begin{eqnarray}\label{sighad}
 \sigma^I(s,q^2) &=& \sum_{a b} \int dx_1 dx_2 f_a (x_1,\mu_F^2) f_b (x_2,\mu_F^2) 
\nonumber\\
&&\times ~ \hat{\sigma}^I_{a b} (\hat{s},q^2,\mu_F^2),
\end{eqnarray}
where $\hat{s} = x_1 x_2 s$, $s$ is the hadronic center of mass energy, and $\hat{\sigma}^I_{a b}$ is the partonic cross section with initial state partons $a$ and $b$.  $I=g$ for Higgs boson production with $q^2=m_H^2$
and $I=q$ for DY production with invariant mass of the dileptons being $q^2$.  $\mu_F$ is
the factorization scale.  
The threshold contribution at the partonic level, denoted by 
$\Delta^{\rm SV}_{I} (z, q^2, \mu_R^2,\mu_F^2)$, normalized by Born cross section $\hat \sigma^{I,(0)}_{ab}$ 
times the Wilson coefficient $C_W^I(\mu_R^2)$,  
is given by 
\begin{equation}\label{sigma}
 \Delta^{\rm SV}_{I} (z, q^2, \mu_R^2, \mu_F^2) = 
{\cal C} \exp ( \Psi^I (z, q^2, \mu_R^2, \mu_F^2, \epsilon )  ) |_{\epsilon = 0} \,,
\end{equation}
where $\mu_R$ is the renormalization scale, the dimensionless variable $z=q^2/\hat s$,
and $\Psi^I (z, q^2, \mu_R^2, \mu_F^2, \epsilon)$ is a finite distribution. 
The symbol ${\cal C}$ implies convolution with the following expansion 
\begin{equation}
 {\cal C} e^{f(z)} = \delta(1-z) + \frac{1}{1!} f(z) + \frac{1}{2!} f(z) \otimes f(z) + \cdots \, .
\end{equation}
Here $\otimes$ means Mellin convolution and $f(z)$ is a distribution of the kind $\delta(1-z)$ 
and ${\cal D}_i$. 
In $d = 4 + \epsilon$ dimensions,
\begin{align*}
&\Psi^I (z, q^2, \mu_R^2, \mu_F^2, \epsilon) =  \Big\{ \ln \Big[ Z^I (\hat{a}_s, \mu_R^2, \mu^2, \epsilon) \Big]^2 \\
& \qquad \quad + \ln \Big|  \hat{F}^I (\hat{a}_s, Q^2, \mu^2, \epsilon)  \Big|^2 \Big\} \delta(1-z) \\
& \qquad \quad + 2 \Phi^I (\hat{a}_s, q^2, \mu^2, z, \epsilon) - 2 {\cal C} \ln \Gamma_{II} (\hat{a}_s, \mu^2, \mu_F^2, z, \epsilon) \,,
\end{align*}
where $\mu$ is the scale introduced to define the dimensionless coupling 
constant $\hat a_s=\hat g_s^2/16 \pi^2$ in dimensional regularization, $Q^2 = - q^2$,
$Z^I (\hat{a}_s, \mu_R^2, \mu^2, \epsilon)$ is the overall operator renormalization constant, which satisfies
\begin{equation*}
 \mu_R^2 \frac{d}{d\mu_R^2} \ln Z^I (\hat{a}_s, \mu_R^2, \mu^2, \epsilon) =  
 \sum_{i=1}^{\infty}  a_s^i (\mu_R^2) \gamma^I_{i-1}\,,
\end{equation*}
where $a_s(\mu_R^2)$ is the renormalized coupling constant that is related to $\hat a_s$ through strong coupling
constant renormalization $Z (a_s(\mu_R^2))$, that is $\hat a_s  = (\mu/\mu_R)^\epsilon 
Z (\mu_R^2) S_\epsilon^{-1} a_s(\mu_R^2)$, $S_\epsilon = \exp\left[(\gamma_E-\ln 4 \pi) \epsilon/2\right]$.
Because of the gauge
and renormalization group invariance, the bare form factors $\hat{F}^I (\hat{a}_s, Q^2, \mu^2, \epsilon)$ satisfy the following differential equation \cite{sudakov} :
\begin{equation*}
 Q^2 \frac{d}{dQ^2} \ln \hat{F}^I  = \frac{1}{2} \Big[ K^I (\hat{a}_s, \frac{\mu_R^2}{\mu^2}, \epsilon ) + G^I (\hat{a}_s, \frac{Q^2}{\mu_R^2}, \frac{\mu_R^2}{\mu^2}, \epsilon ) \Big] \,,
\end{equation*}
where $K^I$ contains all the poles in $\epsilon$ and $G^I$ contains the terms finite in $\epsilon$. Renormalization group invariance of  $\hat{F}^I (\hat{a}_s, Q^2, \mu^2, \epsilon)$ gives
\begin{equation*}
 \mu_R^2 \frac{d}{d\mu_R^2} K^I = - \mu_R^2 \frac{d}{d\mu_R^2} G^I 
= - \sum_{i=1}^{\infty}  a_s^i (\mu_R^2) A^I_i \, .
\end{equation*}
$A^I_i$'s are the cusp anomalous dimensions.  Expanding the $\mu_R^2$ independent part 
of the solution of the RG equation for $G^I$, 
$G^I(a_s(Q^2), 1, \epsilon) = \sum_{i=1}^{\infty} a_s^i(Q^2) G^I_i(\epsilon)$, 
one finds that $G^I_i$ can be decomposed in terms of  
collinear $B^I_i$ and soft $f^I_i$ anomalous dimensions as follows \cite{epsm1}:   
\begin{equation}
 G^I_i (\epsilon) = 2 (B^I_i - \gamma^I_i) + f^I_i + C^I_i + \sum_{k=1}^{\infty} \epsilon^k g_i^{I,k}   \,,
\end{equation}
where $C^I_1 = 0$, $C^I_2 = - 2 \beta_0 g_1^{I,1}$ ,
$C^I_3 = - 2 \beta_1 g_1^{I,1} - 2 \beta_0 ( g_2^{I,1} + 2 \beta_0  g_1^{I,2})$, 
$C^I_4 = -2 \beta_2 g_1^{I,1} -2 \beta_1 (g_2^{I,1} + 4 \beta_0 g_1^{I,2})
-2 \beta_0 (g_3^{I,1}+2 \beta_0 g_2^{I,2}+4 \beta_0^2 g_1^{I,3} )$ 
and $\beta_i$ are the coefficients of the $\beta$ function of strong coupling constant $a_s(\mu_R^2)$,
$\mu_R^2 d a_s(\mu_R^2)/d\mu_R^2 = \epsilon a_s (\mu_R^2) / 2  -\sum_{i=0}^\infty \beta_i a_s^{i+2}(\mu_R^2)$.
The coefficients $g_i^{I,k}$ can be obtained from the form factors. $G^I_1 (\epsilon)$ and $G^I_2 (\epsilon)$ are
known to all orders in $\epsilon$ and $G^I_3 (\epsilon)$ is known to ${\cal O} (\epsilon^3)$ \cite{Gehrmann:2010ue}.

The mass factorization kernel $\Gamma(z,\mu_F^2,\epsilon )$ removes the collinear singularities which arise due to massless partons and it satisfies 
the following RG equation :
\begin{equation}
 \mu_F^2 \frac{d}{d\mu_F^2} \Gamma(z,\mu_F^2,\epsilon) = \frac{1}{2}  P \left(z,\mu_F^2\right) \otimes \Gamma \left(z,\mu_F^2,\epsilon \right) \, .
\end{equation}
$P \left(z,\mu_F^2\right)$ are Altarelli-Parisi splitting functions.  In perturbative QCD,  
$P(z,\mu_F^2)=\sum_{i=1}^\infty a_s^i(\mu_F^2) P^{(i-1)}(z)$.  We find that 
only diagonal elements of the kernel, $\Gamma_{II}(\hat a_s,\mu_F^2,\mu^2,z,\epsilon)$ contribute
to threshold corrections because they contain 
$\delta(1-z)$ and ${\cal D}_0$ at every order perturbation theory while the nondiagornal ones
are regular functions in $z$, that is, $P^{(i)}_{II}(z) = 2 \left[B_{i+1}^I \delta(1-z)+ A_{i+1}^I {\cal D}_0\right] + P_{reg,II}^{(i)}(z)$.

The finiteness of $\Delta_I^{\rm SV}$ demands that the soft distribution function 
$\Phi^{I}(\hat{a}_s, q^2, \mu^2, z, \epsilon)$ also 
satisfies Sudakov-type differential equations \cite{n3losvRavi}, namely,
\begin{equation*}
 q^2 \frac{d}{dq^2} \Phi^I  = \frac{1}{2} \Big[ \overline K^I (\hat{a}_s, \frac{\mu_R^2}{\mu^2}, z,
\epsilon ) + \overline G^I (\hat{a}_s, \frac{q^2}{\mu_R^2}, \frac{\mu_R^2}{\mu^2}, z, \epsilon ) \Big] \, .
\end{equation*}
$\overline{K}^I$ and $\overline{G}^I$ take the forms similar to those of $K^I$ and $G^I$ of the
form factors in such a way that $\Psi^I$ is finite as $\epsilon \rightarrow 0$.  
The solution to the above equation is found to be
\begin{eqnarray}
\Phi^I = \sum_{i=1}^\infty {\hat a}_s^i  \left({q^2 (1-z)^2 \over \mu^2}\right)^{i {\epsilon \over 2}}
S_\epsilon^i \left({i \epsilon \over 1-z}\right) \hat \phi^{I,(i)} (\epsilon)
\end{eqnarray}
where $\hat \phi^{I,(i)}(\epsilon) = \big[\overline K^{I,(i)}(\epsilon) 
+ \overline G^{I,(i)}(\epsilon)\big]/i \epsilon$ and  
$\mu_R^2 d \overline K^I/d\mu_R^2=-\delta(1-z) \mu_R^2 d K^I/d\mu_R^2$.
This implies that $\overline K^{I,(i)} (\epsilon)$ can be 
written in terms of $A^I_i$ and $\beta_i$.  
We define $\overline {\cal G}^I_i(\epsilon)$ through
\begin{eqnarray}
\sum_{i=1}^\infty \hat a_s^i \left( {q_z^2  \over \mu^2}\right)^{i {\epsilon \over 2}}
S_\epsilon^i \overline G^{I,(i)}(\epsilon) =
\sum_{i=1}^\infty a_s^i(q_z^2 ) 
\overline {\cal G}^I_i(\epsilon)
\end{eqnarray}
where $q_z^2=q^2 (1-z)^2$.  Using the fact that $\Delta^{\rm SV}_I$ is finite as $\epsilon \rightarrow 0$, we can express
$\overline {\cal G}^I_i(\epsilon)$ in the form similar to that of $G^I_i(\epsilon)$  
by setting $\gamma^I_i=0, B^I_i=0$ and replacing $f^I_i \rightarrow -f^I_i$ and 
$g^{I,j}_{i} \rightarrow \overline{\cal G}^{I,j}_i$. 
The unknown constants $\overline{\cal G}^{I,j}_i$ can be extracted from the soft part of the partonic 
reactions.  Since $\Phi^I$ results from the soft radiations, the constants $\overline{\cal G}^{I}_i (\epsilon)$ are 
found to be maximally non-Abelian \cite{n3losvRavi}
satisfying 
\begin{eqnarray}
\overline {\cal G}^q_i (\epsilon)= {C_F \over C_A} \overline {\cal G}^g_i(\epsilon)
\label{mna}
\end{eqnarray}
with $C_A = N$, $C_F = (N^2 - 1)/ 2N$, $N$ is the number of colors. Equation~(\ref{mna}) 
implies that the entire soft distribution function for the DY production 
can be obtained from that of Higgs boson production. Substituting ${Z^I}$, the solutions for both ${\hat F}^I$ and ${\Phi}^I$, and $\Gamma_{II}$ in Eq.~(\ref{sigma}),
we obtain $\Delta_I^{\rm SV}$ in powers
of $a_s(\mu_R^2)$ as
\begin{align}
&\Delta_I^{\rm SV}(z) = 
\sum_{i=0}^\infty a_s^i(\mu_R^2) \Delta_{I,i}^{\rm SV} (z,\mu_R^2)\,,  ~~~~ \text{where}
\nonumber\\
&\Delta_{I,i}^{\rm SV} =
\Delta_{I,i}^{\rm SV} (\mu_R^2)|_\delta
\delta(1-z) 
+ \sum_{j=0}^{2i-1} 
\Delta_{I,i}^{\rm SV} (\mu_R^2)|_{{\cal D}_j}
{\cal D}_j \, .
\end{align}
We have set $\mu_R^2=\mu_F^2=q^2$ and their dependence can be retrieved 
using the appropriate renormalization group equation.
$\Delta_{I,i}^{\rm SV}(Q^2)$ are finite and they depend on the anomalous dimensions 
$A^I_i$, $B^I_i$, $f^I_i$ and $\gamma^I_i$, the $\beta$ functions coefficients
$\beta_i$ and $\epsilon$ expansion coefficients of $G^I(\epsilon)$, $g^{I,i}_j$'s and
of the corresponding $\overline {\cal G}^I(\epsilon)$, $\overline{\cal G}^{I,i}_j$'s.  
Up to the two-loop level, all these terms are known to sufficient accuracy
to obtain $\Delta_{I,1}^{\rm SV}$ and $\Delta_{I,2}^{\rm SV}$ exactly. 
At N$^3$LO level, only $\Delta_{I,3}^{\rm SV} |_{{\cal D}_i}$'s were known \cite{n3losv, n3losvRavi} as 
the term $\overline {\cal G}^{g,1}_3$ needed for $\Delta_{I,3}^{\rm SV}|_\delta$ 
has not been available.  Recently 
in Ref.~\cite{Anastasiou:2014vaa}, Anastasiou \textit{et al.} have obtained $\Delta_{g,3}^{\rm SV} |_{\delta}$, using
this we extract $\overline{{\cal G}}^{g, 1}_3$. 
This along with Eq.(\ref{mna}) can be used to determine the corresponding 
$\overline{\cal G}^{q,1}_3$ and hence $\Delta_{q,3}^{\rm SV} |_{\delta}$.
This completes the evaluation of full DY soft plus virtual contributions at N$^3$LO. 
The result for $\overline{\cal G}^{I,1}_3$ is given by 
\allowdisplaybreaks
\begin{widetext}
\begin{align}
&{\overline {\cal G}}^{I,1}_3 = 
C_I \Big\{  {C_A}^2 \Big(\frac{152}{63} \;{\zeta_2}^3 + \frac{1964}{9} \;{\zeta_2}^2
+ \frac{11000}{9} \;{\zeta_2} {\zeta_3} - \frac{765127}{486} \;{\zeta_2}
+\frac{536}{3} \;{\zeta_3}^2 - \frac{59648}{27} \;{\zeta_3} - \frac{1430}{3} \;{\zeta_5}
+\frac{7135981}{8748}\Big)
\nonumber\\
& \quad
+{C_A} {n_f} \
\Big(-\frac{532}{9} \;{\zeta_2}^2 - \frac{1208}{9} \;{\zeta_2} {\zeta_3}
+\frac{105059}{243} \;{\zeta_2} + \frac{45956}{81} \;{\zeta_3} 
+\frac{148}{3} \;{\zeta_5} - \frac{716509}{4374} \Big)
+ {C_F} {n_f} \
\Big(\frac{152}{15} \;{\zeta_2}^2 
\nonumber\\   \label{gb31g}
& \quad
- 88 \;{\zeta_2} {\zeta_3} 
+\frac{605}{6} \;{\zeta_2} + \frac{2536}{27} \;{\zeta_3}
+\frac{112}{3} \;{\zeta_5} - \frac{42727}{324}\Big)
+ {n_f}^2 \
\Big(\frac{32}{9} \;{\zeta_2}^2 - \frac{1996}{81} \;{\zeta_2}
-\frac{2720}{81} \;{\zeta_3} + \frac{11584}{2187}\Big)  \Big\} \,,
\end{align}
%
with $n_f$ being the number of light flavors and $C_I \equiv C_A, C_F $ for $I = g, q$, respectively. The $\Delta_{q,3}^{\rm SV} |_{\delta}$ is given by
\begin{align}
&\Delta^{\rm SV}_{q,3}|_\delta = 
C_A^2 {C_F} \Big(\frac{13264}{315} \;{\zeta_2}^3 + \frac{14611 \
}{135} \;{\zeta_2}^2 - \frac{884}{3} \;{\zeta_2} {\zeta_3} + 843 \
{\zeta_2} - \frac{400}{3} \;{\zeta_3}^2 + \frac{82385}{81} \;{\zeta_3} - 204 \;{\zeta_5} 
- \frac{1505881}{972}\Big) 
\nonumber \\
& \quad
+ {C_A} C_F^2 \
\Big(-\frac{20816}{315} \;{\zeta_2}^3 -\frac{1664}{135} \;{\zeta_2}^2 
+\frac{28736}{9} \;{\zeta_2} {\zeta_3} - \frac{13186}{27} \;{\zeta_2} 
+\frac{3280}{3} \;{\zeta_3}^2 - \frac{20156}{9} \;{\zeta_3} - \frac{39304}{9} \;{\zeta_5} 
+ \frac{74321}{36}\Big) 
\nonumber \\
& \quad
+ {C_A} {C_F} {n_f} \
\Big(-\frac{5756}{135} \;{\zeta_2}^2 + \frac{208}{3} \;{\zeta_2} {\zeta_3} - \frac{28132}{81} \;{\zeta_2}
- \frac{6016}{81} \;{\zeta_3} - 8 \;{\zeta_5} + \frac{110651}{243}\Big)
+ C_F^3 \
\Big(-\frac{184736}{315} \;{\zeta_2}^3 
+ \frac{412}{5} \;{\zeta_2}^2 
\nonumber \\
& \quad
+ 80 \;{\zeta_2} {\zeta_3} 
-\frac{130}{3} \;{\zeta_2} +\frac{10336}{3} \;{\zeta_3}^2 - 460 \;{\zeta_3} 
+ 1328 \;{\zeta_5} - \frac{5599}{6}\Big) 
+ C_F^2 {n_f} \Big(\frac{272}{135} \;{\zeta_2}^2 
- \frac{5504}{9} \;{\zeta_2} {\zeta_3} +\frac{2632}{27} \;{\zeta_2} 
+ \frac{3512}{9} \;{\zeta_3} 
\nonumber\\
& \quad
+ \frac{5536}{9} \;{\zeta_5} 
-\frac{421}{3}\Big)
+{C_F} n_{f,v} \Big( \frac{N^2 -4}{N} \Big)  \
\Big(-\frac{4}{5} \;{\zeta_2}^2 + 20 \;{\zeta_2} + \frac{28}{3} \;{\zeta_3}
-\frac{160}{3} \;{\zeta_5} + 8 \Big) + {C_F} n_f^2 \
\Big(\frac{128}{27} \;{\zeta_2}^2 + \frac{2416}{81} \;{\zeta_2} 
\nonumber \\   \label{cq3delta}
& \quad
- \frac{1264}{81} \;{\zeta_3} - \frac{7081}{243}\Big) \,,
%
\end{align}
%
where, $n_{f,v}$ is proportional to the charge weighted sum of the quark flavors \cite{Gehrmann:2010ue}. 
\begin{table}[h!]
\begin{tabular}{ l  c  c  c  c  c  c  c  c  c  c }
    \hline\hline
     $Q$ (GeV)~~ & 30 & 50 & 70 & 90 & 100 & 200 & 400 & 600 & 800 & 1000 \\
    \hline
     $10^{3} \delta_{\text{N$^3$LO}}$ (nb) & 11.386 & 2.561 & 1.724 & 140.114 & 5.410 & 4.567 $10^{-2}$ & 3.153 $10^{-3}$ & 
                                   6.473 $10^{-4}$ & 2.006 $10^{-4}$ & 7.755 $10^{-5}$ \\
     $10^{3} {\cal D}_{\text{N$^3$LO}}$ (nb) & -8.397 & -2.053 & -1.466 & -124.493 & -4.865  
                                      & -4.421 $10^{-2}$ & -3.368 $10^{-3}$ & 
                                        -7.455 $10^{-4}$ & -2.456 $10^{-4}$ & -9.959 $10^{-5}$ \\
     NNLO (${\rm SV}$) & 0.497 & 0.147 & 0.117 & 10.749 & 0.436 & 4.917 $10^{-3}$ & 4.364 $10^{-4}$ & 1.032 $10^{-4}$ 
                 & 3.538 $10^{-5}$ & 1.480 $10^{-5}$ \\
    NNLO & 0.543 & 0.158 & 0.124 & 11.296 & 0.458 & 5.233 $10^{-3}$ & 4.694 $10^{-4}$ & 1.116 $10^{-4}$ 
         & 3.836 $10^{-5}$ & 1.607 $10^{-5}$ \\
    N$^3$LO $({\rm SV})$ & 0.500 & 0.148 & 0.118 & 10.765 & 0.436 & 4.918 $10^{-3}$ & 4.362 $10^{-4}$ & 1.032 $10^{-4}$ 
                   & 3.534 $10^{-5}$ & 1.478 $10^{-5}$ \\
    N$^3$LO$_{\rm SV}$ & 0.546 & 0.158 & 0.124 & 11.311 & 0.459 & 5.234 $10^{-3}$ & 4.692 $10^{-4}$ & 1.116 $10^{-4}$ 
                   & 3.832 $10^{-5}$ & 1.605 $10^{-5}$ \\
    \hline\hline
    \end{tabular}
 \caption{Contributions of $\delta_{{\rm N}^3{\rm LO}}$, ${\cal D}_{{\rm N}^3{\rm LO}}$, NNLO (SV), exact NNLO, N$^3$LO (SV) and N$^3$LO$_{\rm SV}$}
 \label{table:perc}
\end{table}

 \end{widetext}
We present the contribution from $\Delta^{\rm SV}_{q,3}|_\delta$ to pure N$^3$LO$_{\rm SV}$ as $\delta_{\text{N$^3$LO}}$ 
and the contributions from $\Delta^{\rm SV}_{q,3}|_{{\cal D}_i}$s to pure N$^3$LO$_{\rm SV}$ as ${\cal D}_{\text{N$^3$LO}}$ 
in the Table \ref{table:perc} for different invariant masses ($m_{l^+l^-} \equiv Q$) of the dileptons.
We have used $\sqrt{s} = 14$ TeV for the LHC, number of light quark flavors $n_f = 5$, Fermi constant $G_F = 4541.68$ pb, the $Z$ boson 
mass $m_Z$ = 91.1876 GeV, and 
top quark mass $m_t$ = 173.4 GeV throughout. 
The strong coupling constant $\alpha_s (\mu_R^2)$ is evolved 
using the 4-loop renormalization group equations with 
$\alpha_s^{\text{N$^3$LO}} (m_Z ) = 0.117$ and for parton density sets we use 
Martin-Stirling-Thorne-Watt (MSTW) 2008NNLO \cite{Martin:2009iq}.

\begin{figure}[h]
\includegraphics[width=0.4\textwidth]{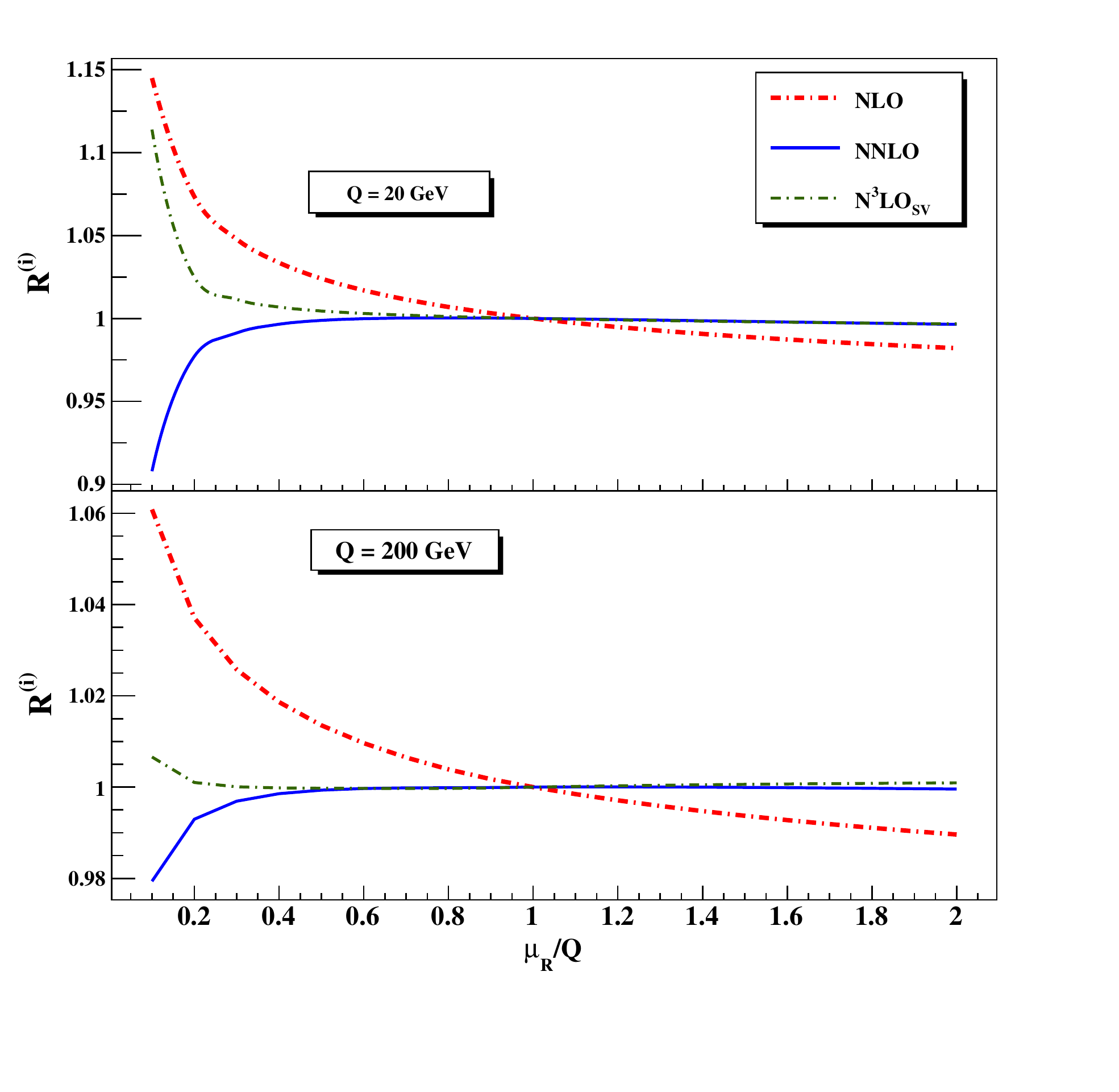}
\vspace{-1cm}
\caption{
\label{fig:plo}
Scale variation.}
\vspace{-0.5cm}
\end{figure}

We find that the $\delta$ contribution is almost equal and opposite in sign to the sum of the contributions
from the ${\cal D}_i$s. Hence, adding the $\delta$ part reduces the pure N$^3$LO$_{\rm SV}$ term to $1$ order in magnitude, 
establishing the dominance of the $\delta$ term. 
We have studied the effect of threshold corrections resulting from
distributions such as $\delta(1-z)$ and ${\cal D}_i$
both at NLO as well as NNLO levels.  In the following, we report our findings
based on the numerical analysis presented in the table 
for two different ranges of $Q$, namely,  $Q=200-900$ GeV (above $m_Z$) 
and $30-110$ GeV (below $m_Z$).  At NLO, if we keep only the distributions and drop
contributions from hard radiations coming from $q \overline q$ and $q(\overline q) g$
initiated processes, we find that the resulting NLO corrected cross section is 
about 95\% of the exact result at NLO level.  Similarly, if we keep the distributions and drop
all the hard radiations both in NLO terms as well as in NNLO terms, 
we find that the resulting NNLO corrected result [NNLO(SV)] is about 95\% of the exact result at 
NNLO level. Hence, it is expected that the sum (N$^3$LO$_{\rm SV}$) of threshold contributions of N$^3$LO terms and the exact NNLO corrected result would constitute the dominant contribution at N$^3$LO level. Like NNLO terms, the threshold contributions in N$^3$LO terms are
also moderate and hence the perturbation theory behaves well.  
In Fig.~\ref{fig:plo}, we have shown the dependence of our result
on the renormalization scale at various orders in perturbation theory.
We have plotted 
$R^{(i)} = \sigma^{i(\mu_R)}/\sigma^{i(Q)}$ where $i =$ NLO, NNLO, N$^3$LO$_{\rm SV}$
versus $\mu_R/Q$ and the reduction in the scale dependence is evident as we increase the
order of perturbation.

%
%

To summarize, we present a systematic way of computing the threshold corrections
to inclusive Higgs boson and DY productions in perturbative QCD.  We have used several 
properties of QCD amplitudes, namely, factorization of soft and collinear divergences,
renormalization group invariance and resummation of threshold contributions. 
For the first time we show that the recent N$^3$LO soft plus virtual contribution to the Higgs boson production cross section
can be used to obtain the corresponding $\delta(1-z)$ part of DY production at N$^3$LO.
We also present numerical results to establish the importance of the $\delta$ term. We find that the impact
of the $\delta$ contribution is quite large to the pure N$^3$LO$_{\rm SV}$
correction.
We have also demonstrated the dominance of threshold corrections at every order
in perturbation theory. We expect that the results presented in this Letter
will not only be a benchmark for a full N$^3$LO contribution but also
an important step in the precision study with the Drell-Yan process at the LHC.
\\ 
%

T.A., M.M. and N.R. thank for the hospitality provided by the Institute of Mathematical Sciences (IMSc)
where the work was carried out. We thank the staff of IMSc computer center for their help. We thank M. K. Mandal for his generous help.
We also thank the referees for useful suggestions.
The work of T.A., M.M. and N.R. has been partially supported by funding from RECAPP, Department of Atomic Energy, Government of India.
\vspace{-1.2cm}

\theendnotes

\vspace{0.2cm}

\end{document}